\documentstyle[11pt,newpasp,twoside,epsfig,epsf]{article}
\def\edcomment#1{\iffalse\marginpar{\raggedright\sl#1\/}\else\relax\fi}
\reversemarginpar

\begin{document}
\title{HST/WFPC2 observations of the core of KjPn 8}
\author{Jos\'e Alberto L\'opez }
\affil{Instituto de Astronom\'{\i}a, Universidad Nacional Aut\'onoma de
M\'exico,  Apartado Postal 877, 22800 Ensenada, B.C., M\'exico}
\author{John Meaburn }
\affil{NRAL, University of Manchester, Jodrell Bank Observatory, Nr Macclesfield, Cheshire, UK, SK11 9DL}
\author{Luis F. Rodr\'{\i}guez }
\affil{Instituto de Astronom\'{\i}a, Universidad Nacional Aut\'onoma de
M\'exico, Apdo. Postal 70-264, M\'exico, D.F.,
04510, M\'exico}
\author{Roberto V\'azquez}
\affil{Instituto de Astronom\'{\i}a, Universidad Nacional Aut\'onoma de
M\'exico,  Apartado Postal 877, 22800 Ensenada, B.C., M\'exico}
\author{Wolfgang Steffen }
\affil{Instituto de Astronom\'{\i}a y Meteorolog\'{\i}a, Universidad de
Guadalajara,  Av. Vallarta 2602, 44130 Guadalajara, Jal., M\'exico }
\author{Myfanwy Bryce}
\affil{NRAL, University of Manchester, Jodrell Bank Observatory, Nr Macclesfield, Cheshire, UK, SK11 9DL}

\begin{abstract}
Narrow-band images of the core of the extraordinary poly-polar planetary
nebula KjPn 8 have been obtained with the WFPC2 camera on board the Hubble 
Space Telescope. Spasmodic bipolar ejections, in changing directions have
occurred over thousands of years to create KjPn 8. The central star is finally
revealed in these observations and its compact nebular core is resolved into
a remarkably young $\approx$ 500 years old, elliptical ionized ring. The 
highest speed bipolar outflows are perpendicular to this central ring which
is identified as the latest event in the creation of this nebula. The 
formation history of KjPn 8 has involved two distinct planetary nebula-like
events, probably originating from a binary core evolution with components of
similar mass.
\end{abstract}

\section{Introduction}
The extraordinary nature of the poly-polar planetary nebula, KjPn~8,  has become
apparent during a series of ground--based observations (L\'opez, V\'azquez, \&
Rodr\'{\i}guez 1995; L\'opez et al. 1997, L\'opez et al. 1999; Steffen \& L\'opez 1998; V\'azquez, Kingsburgh, \& L\'opez 1998). A distance estimation 
to KjPn 8 of 1600 $\pm$ 230 pc, has been deduced by a combination of proper motion and kinematical measurements (Meaburn 1997). It is the 14$^\prime$ $\times$ 4$^\prime$ ~extent of the largest lobes, with knots C$_{1}$--C$_{2}$ 
at their extremities as shown in Fig. 1a, compared with the few arcsec
diameter of the bright nebular core, that first indicated the unusual nature 
of this nebula.
Also, secondary, smaller lobes, delineated by knots A$_{1}$--A$_{2}$  in 
Figure 1 (top), have a distinctly different axis from C$_{1}$--C$_{2}$. 

The recognition of
multiple outflows along different axes lead originally to their
interpretation as the action of a bipolar, rotating, episodic jet or BRET
(L\'opez et al. 1995). 

However, it has required optical imagery with the Hubble
Space Telescope, to reveal the elliptical ionized ring which constitutes the
nebular core of KjPn 8 and to locate the central star which is now shown to be
at the center of this ionized central ring. This ring is clearly the ionized,
inside surface of a 7 arcsec diameter ring of excited H$_{2}$ (L\'opez et al.
1999), see Figure 1 (bottom), itself located within a central 30 arcsec 
diameter CO disk (Foreveille et al. 1998) whose axis is aligned with the 
bipolar outflows defined by A$_{1}$--A$_{2}$ in Figure 1 (top). 



\section{Dynamical characteristics}

The largest lobes, with knots C$_{1}$--C$_{2}$ at their extremities, 
from their linear dimensions (6.5 $\times$ 1.9 pc) and kinematics, must be 
$1 - 2 ~\times$ 10$^{4}$ yr old (Steffen \& L\'opez 1998), whereas the knots, 
A$_{1}$--A$_{2}$, aligned with the axis of the ring have a 
kinematical age $\leq 3,400$ yr as given directly by their angular 
displacements from the nebular core combined with measurements of their 
expansion proper motions (Meaburn 1997). This particular timescale estimation 
is  independent of the distance to KjPn~8. The ionized ring itself, if
expanding at a constant 40 km s$^{-1}$, would only take $\sim 500$ yr to reach its present 2.7 arcsec ($\equiv$ 0.02 pc) radius. 

These temporal differences indicate that the A$_{1}$--A$_{2}$ high-velocity (320 km s$^{-1}$) knots
and associated  bipolar outflows were formed prior to the present central
ionized ring. Furthermore, the largest features, culminating in the knots  C$_{1}$--C$_{2}$, must have formed along a different ejection axis well before any of the central ionized and molecular circumstellar structures 
had been formed. 

The small dimensions of the central ionized
ring, associated molecular material and low excitation nebular spectrum
(V\'azquez et al. 1998), indicate that the physical characteristics of the 
core of KjPn 8 are representative of a very young PN. Moreover, its ionic abundances, with enhanced He and N, (V\'azquez et al. 1998) correspond to
extreme type I PNe that are identified with massive ($> 2.4 M_{\odot}$)            progenitors (Peimbert \& Torres-Peimbert 1994) which should evolve
relatively quickly during the PN stage towards higher effective temperatures
and consequently higher excitation conditions. This implies that the core of
KjPn 8 has only reached photoionization conditions during the last few hundred
years and the formation of the bipolar high-speed (A$_{1}$--A$_{2}$) outflows
ocurred shortly before, during the pre-planetary nebula stage.

\begin{figure}
\plotone{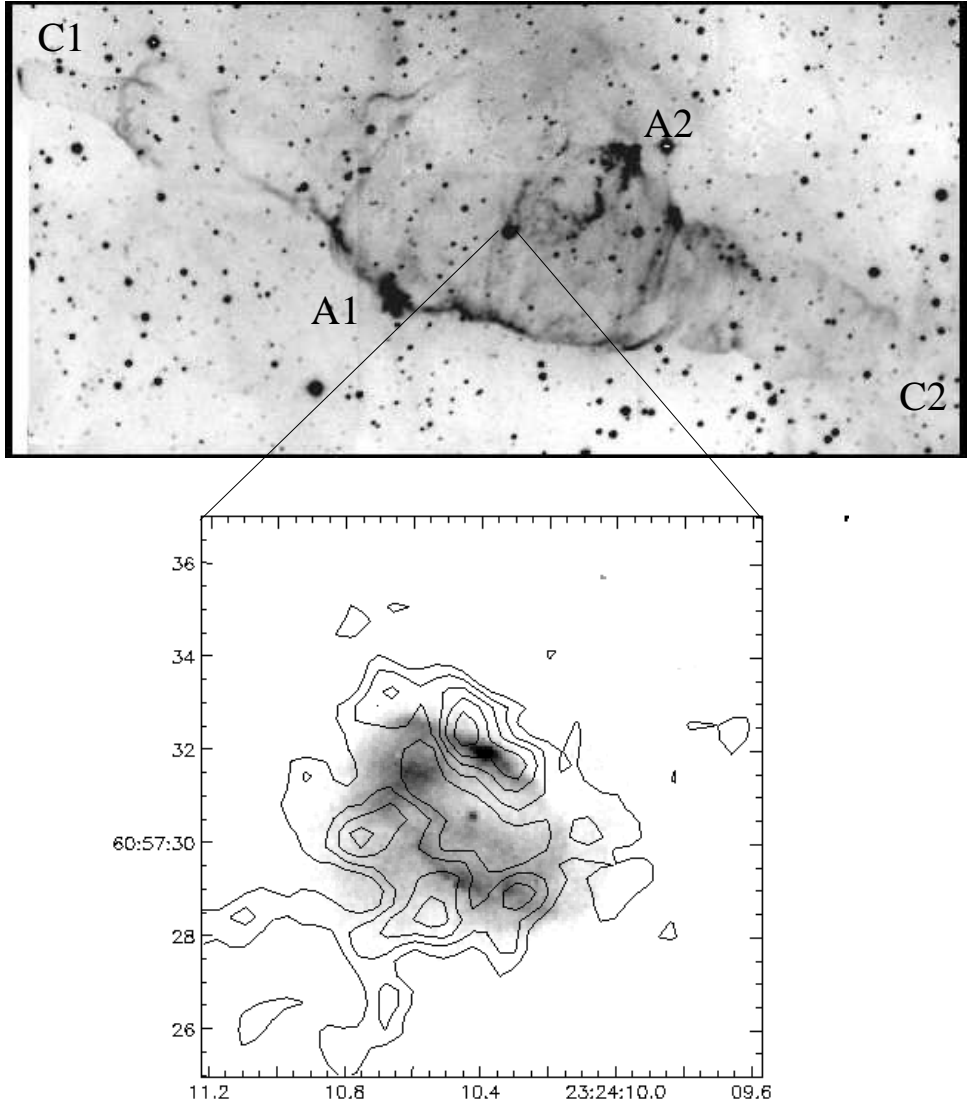}
\caption{Top- A deep H$\alpha$ wide field, ground-based, image of the
polypolar nebula KjPn 8. The symmetric knots C$_1$--C$_2$ (PA 72$^\circ$ ) and A$_1$--A$_2$ (PA 126$^\circ$ ) are located at the tips of independent
bipolar outflows. The whole nebula is 14$^\prime$ $\times$ 4$^\prime$ ~in extent. Bottom- A contour map of excited H$_2$ $\upsilon = 1 \rightarrow 0$ S(1) at 2.122 $\mu$m is shown overlayed on the [S~II] HST image of the core of KjPn 8 where the central star is apparent. The H$_2$ emission surrounds and shares the
morphology and orientation of the ionized ring, as also does a larger disk of
CO (Forveille et al. 1998) not shown here, confirming a second heavy mass-loss
episode in KjPn 8.}

\end{figure}

For the current core conditions, the associated CO and H$_2$ 
molecular material must be related to a second heavy mass-loss episode prior
to the formation
of the ionized nebular core. The disk-like structure and 
common orientations of the molecular material and ionized nebular ring  
confirm their connection in this second event. These characteristics are 
incompatible with the expected conditions that the core must have had at the 
time when the C$_{1}$--C$_{2}$ bipolar outflows where triggered.
The arguments lead to  the conclusion that the formation of the giant 
bipolar envelope had its origin in a different event, unrelated to the 
creation of the present nebular core and associated bipolar outflows. 

KjPn 8 thus unfolds a unique situation among PNe: the creation of a double
planetary nebulae event. A large bi-conical nebula was formed through
episodic jets (Steffen \& L\'opez 1997) along PA 72$^\circ$ (C$_1$-C$_2$),
1 - 2 $\times 10^4$ years ago. These jets have now ceased their activity. A second planetary nebula event is initiated $\leq 3,400$ years ago ejecting high-velocity bipolar outflows along PA 126$^\circ$ (A$_1$-A$_2$). Unequivocal
signatures of a second heavy mass-loss, superwind episode accompany 
this second event. Massive molecular disks of  CO and H$_2$  surround
now a very young ionized ring, all of which are perpendicular to the most 
recent A$_1$-A$_2$ bipolar outflows.

A possible explanation for the formation of  KjPn 8 is that here we are
witnessing the near-simultaneous death of two relatively massive stars in a
binary system either with a separation large enough for no  effective mass
transfer to take place (separations from several tens to a few  hundred
astronomical units) or detached binaries (with separations of the order of
a few tens of AU) where the evolution of an originally less massive secondary
may be speeded up by wind accretion from the primary so both reach the PN 
stage one shortly after the other, within 1-2 $\times 10^4$ years. 
Two PNe-type events have thus  been consecutively produced from a binary core
where the influence of the companion has probably aided in the production of
bipolar outflows on each occasion and for each event having its own symmetry axis. Full details of this work will appear elsewhere.

\end{document}